\documentclass[acmsmall]{acmart}

\def\BibTeX{{\rm B\kern-.05em{\sc i\kern-.025em b}\kern-.08emT\kern-.1667em\lower.7ex\hbox{E}\kern-.125emX}}
    
\usepackage{li-stdpkgs}
\usepackage{li-typography}

\begin{document}

%
\title{\BenchName: A Parallel Sparse Tensor Algorithm Benchmark Suite}

%
\author{Jiajia Li}
\email{jiajia.li@pnnl.gov}
\affiliation{%
  \institution{Pacific Northwest National Laboratory}
  \streetaddress{902 Battelle Blvd}
  \city{Richland}
  \state{WA}
  \country{USA}
  \postcode{99354}
}

\author{Yuchen Ma}
\email{m13253@hotmail.com}
\affiliation{%
  \institution{Hangzhou Dianzi University}
  \city{Hangzhou}
  \country{China}
  \postcode{310005}
}

\author{Xiaolong Wu}
\email{xlwu@vt.edu}
\affiliation{%
  \institution{Virginia Tech}
  \city{Blacksburg}
  \state{VA}
  \country{USA}
  \postcode{24061}
}

\author{Ang Li}
\email{ang.li@pnnl.gov}
\affiliation{%
  \institution{Pacific Northwest National Laboratory}
  \city{Richland}
  \state{WA}
  \country{USA}
  \postcode{99354}
}

\author{Kevin Barker}
\affiliation{%
 \institution{Pacific Northwest National Laboratory}
  \city{Richland}
  \state{WA}
  \country{USA}
  \postcode{99354}
}

%
\renewcommand{\shortauthors}{J. Li et al.}

%
\begin{abstract}
Tensor methods have gained increasingly attention from various applications, including machine learning, quantum chemistry, healthcare analytics, social network analysis, data mining, and signal processing, to name a few.
Sparse tensors and their algorithms become critical to further improve the performance of these methods and enhance the interpretability of their output.
This work presents a sparse tensor algorithm benchmark suite (\BenchName) for single- and multi-core CPUs. 
To the best of our knowledge, this is the first benchmark suite for sparse tensor world.
\BenchName targets on: 1) helping application users to evaluate different computer systems using its representative computational workloads; 2) providing insights to better utilize existed computer architecture and systems and inspiration for the future design. 
This benchmark suite will be publicly released.

\end{abstract}

\maketitle

\section{Introduction}  \label{sec:intro}
Tensors draw increasing attention from various domains, such as machine learning, quantum chemistry, healthcare analytics, social network analysis, data mining, and signal processing, to name a few.
Tensor methods have been noted for their ability to discover multi-dimensional inherent relationships from underlying application logic. 
A tensor is a multi-dimensional array, generalized matrices and vectors to more dimensions.
In data-oriented tensor applications ~\cite{Ho:2014:marble,Henderson:2017:Granite,Papalexakis:2011:tensor-app,Chi:2012:cp-apr,Sidiropoulos:2017:survey}, sparse tensors are often found, where most of its entries are zeros.

High-performance computing (HPC) now enters the era of extreme heterogeneity. As many general purpose accelerators, such as Graphics Processing Unit (GPUs), Intel Xeon Phi, and Field-Programmable Gate Array (FPGAs), and domain-specific architectures, such as near-memory, thread migratory architecture Emu~\cite{Hein:2018:emu} and Google Tensor processing unit (TPU)~\cite{Jouppi:2017:TPU}, emerge, it is natural to ask whether the critical sparse-tensor based algorithms can
 be efficiently executed on these platforms, with their non-regular parallelism to be effectively exploited. However, the lack of a concrete, comprehensive, and easy to use sparse tensor algorithm benchmark suite prevents us from answering this question easily.


In this paper, we fill this gap by proposing a PArallel Sparse Tensor Algorithm benchmark suite called \BenchName. \BenchName incorporates various sparse tensor algorithms and operations, serving as a handy tool for application developers to assess different platforms, in terms of their tensor
 processing capability. 
Consisting state-of-the-art sequential and parallel versions, while adopting the most popular sparse tensor format COO, \BenchName can also supply a fair baseline for evaluating performance improvement brought by new sparse tensor methods. Application developers seeking to exploit tensor
 sparsity for further performance speedup may also find it useful as a good reference. 


This paper makes the following contributions:
\begin{itemize}
	\item We show the importance of sparse tensor operations and tensor methods in diverse tensor applications. (\Cref{sec:methods-apps})
	\item We extract 12 computational sparse tensor operations as \BenchName workloads: Tensor Element-Wise operations -- \TEW-eq (addition/subtraction/multiplication/division) and \TEW (addition/subtraction/multiplication), Tensor-Scalar operations -- \TS addition/multiplication, Tensor-Times-Vector operation (\TTV), Tensor-Times-Matrix operation (\TTM), and Matricized Tensor Times Khatri-Rao Product (\MTTKRP). (\Cref{sec:workload})
	\item We implement sequential and multicore parallel algorithms for all workloads, based on the most popular coordinate (COO) sparse tensor format. Our experiments and analysis show the usefulness of \BenchName on single- and multi-core CPUs. (\Cref{sec:impl}, \ref{sec:dataset}, \ref{sec:exp})
\end{itemize}

\section{Motivation}  \label{sec:motiv}
This work is motivated by first demonstrating the challenges of sparse tensor algorithms and then illustrating that existed libraries or toolsets cannot meet the requirements of a benchmark suite from diversity, timeliness, research support, and dataset four aspects.

\subsection{Challenges of Sparse Tensor Algorithms}
We summarize the challenges of sparse tensor algorithms into five points:

\emph{The curse of dimensionality} refers to the issue that the number of entries of an intermediate or output tensor can grow exponentially with the tensor order, resulting in significant computational and storage overheads.
Even when the tensor is structurally sparse, meaning it consists mostly of zero entries, the execution time of one important tensor method, CP decomposition introduced in \Cref{sec:methods}, generally grows quadratically with the number of non-zeros~\cite{Bader:2007:tensortoolbox-sparse,Bader:2017:tensortoolbox-pak}.
And there is an increasing interest in applications involving a large number of dimensions~\cite{Novikov:2015:tnn,DeLathauwer:2017:tensorize,Lebedev:2014:cnn-cp}, which makes this problem more difficult.

\emph{Mode orientation} refers to the issue of a particular storage format favoring the iteration of tensor modes in a certain sequence, which is of particular concern in the sparse case.
Since most methods of interest require more than one sequence, being efficient for every sequence generally requires storing the tensor in multiple formats, thereby trading extra memory for speed.
A question arises, that is whether one can achieve both a neutral mode orientation and compact storage which also helps reduce memory footprint.

\emph{Tensor transformation(s)} refers to a common pattern for attaining speed in some implementations of tensor algorithms, which starts by reorganizing the tensor into a matrix and then perform equivalent matrix operations using highly tuned linear algebra libraries.
Done na{\"i}vely, this approach appears to require an extra memory copy, which can even come to dominate the overall running time.
We observe instances in which such a copy consumes 70\% or more of the total running time (in the case of a \TTM operation). 

\emph{Irregularity} refers to two issues. The first is that a tensor may have dimension sizes that vary widely; the second is that a sparse tensor may have an irregular non-zero pattern, resulting in irregular memory references.

\emph{Arbitrary tensor orders} generate various implementations of a tensor operation.
For the sake of performance, programmers usually implement and optimize third-order tensor algorithms apart from higher-order ones.
These implementations makes no one optimization method can fit all variations, e.g., different number of loops and diverse memory access behavior. 

These challenges bring non-trivial computational and storage overheads, and some of them are even harder to overcome than their counterparts in classical linear algebra.
To overcome these challenges, it is necessary to build a sparse tensor benchmark suite to evaluate diverse algorithms and computer systems.

\subsection{Requirements for a Benchmark Suite}
By surveying some benchmark suites~\cite{dixit1991spec,wang2014BigDataBench,che2009Rodinia,kleinosowski2002minnespec,lee1997mediabench,poovey2009benchmark,bienia2008parsec}, we present the following four requirements for a benchmark suite.

\emph{Diversity.}
We analyze diversity from two aspects: \emph{application diversity} and \emph{platform diversity}.
\emph{Application diversity} means a benchmark suite should represent a broad and representative applications. For example, EEMBC benchmark suite~\cite{poovey2009benchmark} is developed for autonomous driving, mobile imaging, the Internet of Things, mobile devices, and many other applications; PARSEC benchmark suite~\cite{bienia2008parsec} covers computer vision, video encoding, financial analytics, animation physics and image processing, etc..
Sparse tensor methods have a broad application domains (refer to ~\Cref{sec:apps}), the workloads in our benchmark suite also need to represent the diversity of these domains.
\emph{Platform diversity} is that a benchmark suite should support different computer architectures and platforms, especially the emerging ones.
For example, SPEC benchmarks~\cite{dixit1991spec} supports scientific applications on diverse platforms: CPUs, distributed platforms, accelerators, web servers, cloud platforms, etc. A recent Tartan benchmark~\cite{li2018tartan} collected kernels from machine learning, data analysis, high performance simulation, molecular dynamics and so on and optimized them on multi-GPU platforms.

\emph{Timeliness.}
A benchmark suite should be kept updated by including the state-of-the-art \emph{data structures}, \emph{algorithms}, and \emph{optimization techniques}.
Especially for sparse data, the data structure is closely relevant to the performance of its algorithm. This phenomenon has been observed from sparse matrices, where different sparse formats behave quite differently on diverse input matrices~\cite{Zhao:2018:spmv,Li:2013:smat,sedaghati2015spmv,su2012clspmv}.
As mentioned in the work~\cite{bienia2008parsec}, an outdated algorithm cannot well reflect the current status of an application. This can easily mislead the researchers using this benchmark suite to test a machine's behavior.
As the computer architectures keep evolving, an under-optimized code, e.g., sequential benchmark programs for a multicore machine, cannot be a fair measurement. 
Optimized implementations for architectures have to be taken account.

\emph{Research support.}
Research support also includes two aspects: support of domain research and benchmarked workload research.
The former requires a benchmark suite to be \emph{compatible}, while the latter requires it to be \emph{extensible}. 
Since some workloads are still open research problems in an application domain, a compatible workload should be able to do easy comparison with other research work by supporting unified input/output format and interface to high-level applications.
The workload research mainly develops its high performance, power or other efficiency. An extensible workload is easy to be assembled with new data structures, algorithms, and optimization techniques.

\emph{Dataset.}
Data becomes essential to data-intensive applications and their workloads which widely exist in real world. Traditionally, two types of dataset are considered:\emph{synthetic} and \emph{real} data. Real data comes directly from real-world applications, which can best reflects the application features. However, due to some factors such as information protection, sensitive data, etc., researchers are usually short of data. Thus, synthetic data are generated according to some regulations and scenarios from applications.

\subsection{\BenchName in Need}
Some tensor libraries or toolsets have existed for sparse tensor algorithms.
The most popular libraries are Tensor Toolbox~\cite{Bader:2017:tensortoolbox-pak} and TensorLab~\cite{Vervliet:2016:tensorlab-pak}. They are both implemented using MATLAB. The main shortcoming is that these two libraries are hard to be implemented on various platforms, such as multicore CPUs and GPUs, which violates the platform diversity requirement. Besides, their performance efficiency is low because of MATLAB environment.
Recently, many other highly performance efficient libraries emerge, such as SPLATT~\cite{Smith:2015:splatt}, Cyclops Tensor Framework (CTF)~\cite{Solomonik:2015:sparse-ctf}, DFacTo~\cite{Choi:2014:dfacto}, GigaTensor~\cite{Kang:2012:gigatensor}, HyperTensor~\cite{Kaya:2015:dist-cp}, GenTen~\cite{phipps2018genten}, to name a few. However, these libraries are specific to one or two particular sparse tensor operations, this violates the application diversity requirement.
Beyond these, the requirements of timeliness, research support, and dataset are barely met by these libraries.
Our \BenchName is proposed to meet all the requirements from our continuous effort.

\section{Tensor Methods and Applications}  \label{sec:methods-apps}

\begin{table}
	\small
	\begin{tabular}{|l l l|}
		\hline
		\textbf{Domains} & \textbf{Tensor Methods} & \textbf{Workloads} \\ \hline
		 Machine Learning & \CPD, \TPM, Tucker, \TT, hTucker & \TS, \MTTKRP, \TTV, \TTM, \TTT \\ \hline
		 Healthcare Analytics & \CPD & \MTTKRP \\ \hline
		 Social Network Analysis & \CPD, Tucker & \TTM \\ \hline
		 Quantum Chemistry & \CPD, Tucker & \TS, \TEW, \TTM, \MTTKRP, \TTT \\ \hline
		 Brain Signal Analysis & \CPD & \MTTKRP \\ \hline
		 Personalized web search & \CPD, Tucker & \MTTKRP, \TTM \\ \hline
		 Recommendation systems & \CPD, Tucker & \MTTKRP, \TTM \\ \hline
		 Signal Processing & \CPD & \MTTKRP \\ \hline
		 Direct Numerical Simulation & Tucker & \TTM \\ \hline
		 Power Grid & \CPD, Tucker & \MTTKRP, \TTM \\ \hline

	\end{tabular}
	\caption{The relationship between tensor domains, tensor methods, and workloads.}
	\label{tab:apps}
\end{table}

This section describes the broad applications of tensors methods in diverse domains, along with the tensor methods and their computational operations. The summarized form is presented in \Cref{tab:apps}.

\subsection{Tensor Methods}
\label{sec:methods}
In this section, we summarize tensor methods in three categories: tensor decompositions, tensor network models, and tensor regression.
Though tensor network models also belong to tensor decomposition methods, because of their network format and more emphasizing on high-order tensors, we discuss them separately. 

\subsubsection{Tensor Decompositions} 
We introduce three low-rank tensor decompositions which have applications for sparse data.

\textbf{\CPD.}
The CP decomposition (\CPD) was first introduced in 1927 by Hitchcock~\cite{Hitchcock:1927:cp}, and independently introduced by others~\cite{Carroll:1970:cp,Harshman:1970:parafac}. 
\CPD decomposes an $N$th-order tensor into a sum of component rank-one tensors with different weights~\cite{Kolda:2009:survey}.
In a low-rank approximation, a tensor rank $R$ is chosen to be a small number less than $100$.
From a data science standpoint, the results can be interpreted by viewing the tensor as being composed of $R$ latent rank-1 factors.
\CPD has proven both scalable and effective in many applications in ~\Cref{sec:apps}.

Other variants of \CPD exist by restructuring of the factors or their constraints to accommodate diverse situations, such as INDSCAL~\cite{Carroll:1970:cp}, CANDELINC~\cite{Carroll:1980:candelinc}, PARAFAC2~\cite{Harshman:1972:parafac2,Perros:2017:spartan}, and DEDICOM~\cite{Harshman:1970:parafac}.
Many \CPD methods have been proposed in a broad area of research, such as Alternating Least Squares (ALS) based methods~\cite{Harshman:1970:parafac,Kolda:2009:survey,Uschmajew:2016,Kaya:2015:dist-cp}, block coordinate descent (BCD) based methods~\cite{Mohlenkamp:2010:MusingsOM,Li:2015:bcd}, Gradient Descent based methods~\cite{Sorber:2013,Ravindran:2014,Smith:2016:exploration,Beutel:2013:flexifact}, quasi-Newton and Nonlinear Least Squares (NLS) based methods~\cite{Chi:2012:cp-apr,Sorber:2013,Wright:1999:numerical,Tomasi:2006:nls,Ishteva:2011:nls,Savas:2010:nls,Hansen:2015:NewtonbasedOF}, alternating optimization (AO) with the alternating direction method of multipliers (ADMM) based methods~\cite{Boyd:2011:ao-admm,Smith:2017:constrained}, exact line search based methods~\cite{Rajih:2005:els,Sorber:2016:ELP}, and randomized/sketching methods~\cite{Battaglino:2018:randCP,Vervliet:2016:randCP,Papalexakis:2012:parcube,Dehua:2016:spals,Reynolds:2016,Song:2016:STO}.
Sparse \CPD comes from two aspects: the sparse tensor from applications~\cite{Chi:2012:cp-apr,Sidiropoulos:2017:survey,Kolda:2009:survey,Bader:2007:tensortoolbox-sparse,Choi:2014:dfacto,Smith:2015:splatt,Kang:2012:gigatensor,Kaya:2018:dimtree-sparse-cp,Smith:2017:knl,Smith:2016:medium,Li:2018:hicoo,Li:2017:adatm,Ravindran:2014,Li:2018:thesis,phipps2018genten,Choi:2018,Liu:2017:fcoo} and the constrained sparse factors from some \CPD models~\cite{Ho:2014:marble,Henderson:2017:Granite,Papalexakis:2011:tensor-app}.

The computational bottleneck of \CPD is the matriced tensor-times-Khatri-Rao product (\MTTKRP) (will be described in \Cref{sec:mttkrp}).


\textbf{Tucker.}
Tucker decomposition, first introduced by Ledyard R. Tucker~\cite{Tucker:1966}, provides a more general decomposition.
It decomposes an $N$th-order tensor into a small-sized $N$th-order core tensor along with $N$ factor matrices that are all orthogonal.
The core tensor models a potentially complex pattern of mutual interaction between tensor modes.
Its size determined by $N$ ranks which can be chosen according to the work~\cite{Kiers:2003}.
In a low-rank approximation, the rank sizes are usually less than $100$.

Some variants of Tucker decomposition are PARATUCK2~\cite{Richard:1996:cp-tucker}, lossy Tucker decomposition~\cite{Zhou:2014:DecompositionOB}, and so on.
Methods for Tucker decomposition include higher-order SVD (HOSVD)~\cite{DeLathauwer:2000:multi-svd}, truncated HOSVD~\cite{DeLathauwer:2000:multi-svd}, Alternating Least Squares (ALS) based methods~\cite{Kapteyn:1986}, the popular higher-order orthogonal iteration (HOOI)~\cite{DeLathauwer:2000:Tucker-hooi}, Newton–Grassmann optimization~\cite{Elden:2009}.
Sparse Tucker also comes from two aspects: the sparse tensor from applications~\cite{Ma:2018:sptucker-gpu,Li:2018:thesis,Smith:2017:csf-tucker,Liu:2017:fcoo} and the constrained sparse factors.

The computational tensor kernel of Tucker decomposition is the Tensor-Times-Matrix operation (\TTM) (will be described in \Cref{sec:ttm}).

\textbf{\TPM.}
Tensor power method~\cite{DeLathauwer:2000:Tucker-hooi,Anandkumar:2014:survey} is an approach for orthogonal tensor decomposition, which decomposes a symmetric tensor into a collection of orthogonal vectors with corresponding positive scalars as weights. 
Some variations have been proposed~\cite{Anandkumar:2014:survey,Yu:2017}. When the tensor is sparse, we need to use sparse method correspondingly.

The computational tensor kernel of tensor power method is the Tensor-Times-Vector operation (\TTV) (will be described in \Cref{sec:ttv}).

\subsubsection{Tensor Network Models}
\CPD and Tucker decompositions assume a model in which all modes interact with all the other modes, which ignores the situations where modes could interact in subgroups or hierarchies.
Tensor network models decompose a tensor in tensor networks which expose more localized relationships between modes.
Tensor networks have flexibility in modeling and compute/storage efficiency especially for high-order tensors.

\textbf{\TT.}
Tensor Train (\TT) decomposition, also called Matrix Product State (MPS) in quantum physics community~\cite{Cichocki:2016:survey,Grasedyck:2013:survey}, was first proposed by Ivan Oseledets in the work~\cite{Oseledets:2011:tt}.
\TT decomposes a high-order tensor into a linear sequence of tensor-times-tensor/matrix products.
The contraction modes are in small rank sizes in low-rank approximation.

The variants of \TT include tensor chain (\TC), tensor networks with cycles: Projected Entangled Pair States (PEPS)~\cite{Orus:2014:practical}, Projected Entangled Pair Operators (PEPO)~\cite{Evenbly:2009:algorithms}, Honey--Comb Lattice (HCL)~\cite{Giovannetti:2008:quantum}, Multi-scale Entanglement Renormalization Ansatz (MERA)~\cite{Orus:2014:practical}.

The computational tensor kernels of \TT are the Tensor-Scalar (\TS), Tensor-Times-Matrix (\TTM) and Tensor-Times-Tensor (\TTT) operations. \TS and \TTM will be described in \Cref{sec:ts} and \ref{sec:ttm} respectively, and \TTT will be one of our future work. 

\textbf{hTucker.}
Hierarchical Tucker (hTucker) decomposition, also called hierarchical tensor representation, was introduced in~\cite{Hackbusch:2009:htucker,Grasedyck:2010:ht,Grasedyck:2013:survey,Cichocki:2016:survey}.
hTucker recursively splits the set of tensor modes, resulting a binary tree containing a subset of modes at each node.
This binary tree is called dimension tree, and the modes from different nodes do not overlap.
\TT decomposition is a special case of hTucker while the dimension tree is linear and extremely unbalanced.

Variants of hTucker include the Tree Tensor Network States (TTNS) model~\cite{Nakatani:2013:TTNS}, multilayer multi-configuration time-dependent Hartree method (ML-MCTDH)~\cite{Wang:2003}.
Sparsity has been considered by Perros et al. in the work~\cite{Perros:2015:SHT}.

The computational tensor kernels of hTucker are the Tensor-Scalar (\TS), Tensor-Times-Matrix (\TTM) and Tensor-Times-Tensor (\TTT) operations. \TS and \TTM will be described in \Cref{sec:ts} and \ref{sec:ttm} respectively, and \TTT will be one of our future work.

\subsubsection{Tensor Regression}
Tensor regression is an extension of classical regression model, but using tensors to represent input and covariates data.
Tensor regression approximates coefficient tensor with a low-rank decomposition, thus tensor decomposition methods introduced above can be easily adopted here.
Some tensor regression methods have been proposed~\cite{Zhao:2011,Zhou:2013:survey,Romera-Paredes:2013:MML,Wimalawarne:2014,Signoretto:2014,Yu:2016,Yu:2017}.

\subsection{Tensor Applications}
\label{sec:apps}
Tensor methods can be used in applications to expose the inherent relationship in the observed data and to represent the data in a more compressed way.
This section does not keen to give a thorough survey of tensor applications but emphasizes on showing the broad application scenarios tensor methods can be applied and useful in.
Please refer to these surveys for more complete tensor applications~\cite{Kolda:2009:survey,Cichocki:2016:survey,Cichocki:2014:survey,DeLathauwer:2008:block-survey-1,Cichocki:2015:survey,Anandkumar:2014:survey,Sidiropoulos:2017:survey}.

\subsubsection{Machine Learning}

The diversity needs of machine learning algorithms have promoted the exploitation of various tensor-based decompositions, regressions, and techniques from this community.
In particular, the latent variable model, where hidden factors are assumed to express structure in observed data, has been frequently expressed using \CPD~\cite{Huang:2014:tensor-app}, tensor power method~\cite{Anandkumar:2014:survey}, hTucker~\cite{Song:2013:tensor-app}, and other formats~\cite{Ishteva:2013:tensor-app}. 

\CPD, Tucker and \TT decompositions have been leveraged in the context of neural networks~\cite{Novikov:2015:tnn,Janzamin:2015:nn,Yu:2012:tnn,Setiawan:2015:tn,Socher:2013:tn,Hutchinson:2013:tdsn,Lebedev:2014:cnn-cp,Novikov:2018:t3f,Yu:2018:longterm},
with the weight matrix of a fully-connected layer or a convolutional layer stored compressedly in a low-rank tensor, thus reducing redundancies in the network parameterization.
As concerns improving theoretical aspects and understanding of deep neural networks through tensors, Cohen et al.~\cite{Cohen:2015:dl-vs-td} analyzed the expressive power of deep architectures by drawing analogies between shallow networks and the rank-1 \CPD, as well as between deep networks and the hTucker decomposition.
Novikov et al. applied \TT in Google's TensorFlow~\cite{Novikov:2018:t3f,Abadi:2015:tensorflow} which expresses a wide variety of algorithms as operators (graph nodes) that communicate tensor objects through the graph's edges.

Other Machine Learning applications include using \TT to improve Markov Random Field (MRF) inference problem~\cite{Novikov:2014} and extending standard
Machine Learning algorithms such as Support Vector Machines and Fisher discriminant analysis to handle tensor-based input~\cite{Tao:2007:supervised}.
Tensor methods involving other machine learning tasks such as feature selection and multi-way clustering will be discussed in other applications below.


\subsubsection{Healthcare Analytics}

The work on tensor-based healthcare data analysis has been driven by the need of improving the interpretability and the robustness of underlying methods, with the goal that healthcare professionals may eventually use consulting tools based on these methods. 
As a result, recent work has focused on modifying traditional tensor methods like \CPD by adding constraints that better describe the underlying data and exploit domain knowledge. 
One particular focus is handling \emph{sparsity}, which is particularly important when handling
event-recording tensors describing healthcare data~\cite{Ho:2014:limestone,Ho:2014:ntf-app,Ho:2014:marble,Wang:2015:rubik,Perros:2015:SHT,Matsubara:2014:funnel-tensor-app,Zhou:2013:survey}.

\subsubsection{Social Network Analysis}

Some studies have been done on DBLP authorship data~\cite{Papalexakis:2013:moreviewsgraph} by using dynamic/static tensor analysis (include \CPD, Tucker decompositions and their variants) to demonstrate clustering~\cite{Sun:2006:dynamic,Kolda:2008:mettm}, find interesting events (or anomalies) in the users' social activities~\cite{Papalexakis:2012:parcube,Papalexakis:2015:parcube-2}.
Jiang et al. identified patterns in human behavior through a dynamic tensor decomposition of user interactions within a microblogging service~\cite{Jiang:2014:fema-tensor-app}.
Sun et al. demonstrated a sampling-based Tucker decomposition~\cite{Sun:2009:multivis}, to jointly model the sender-recipient interaction and share content within business networks.
The work in~\cite{Benson:2015:tensor-app} utilizes tensors to model higher-order structures,
such as cycles or feed-forward loops in a graph clustering framework.


\subsubsection{Quantum Chemistry}
Tensors have a long history in quantum chemistry because of the nature of high-dimensional data there~\cite{khoromskaia2018tensor}.
Hartree--Fock (HF) is a method of approximation for the energy of a quantum many-body system and large-scale electronic structure calculations.
Koppl et al. proposed sparsity using local density fitting in Hartree--Fock calculations, which heavily involves \TTT and \TTM operations~\cite{Koppl:2016}.
Lewis et al. introduced a clustered low-rank tensor format to exploit element and rank sparsities~\cite{Lewis:2016}.
Block sparsity has been utilized in coupled-cluster singles and doubles (CCSD) in the work~\cite{Peng:2016,Calvin:2016:tiledarray-pak,Kaliman:2017,Epifanovsky:2013,Manzer:2017}.
Scaled opposite spin second order Møller-Plesset perturbation theory (SOS-MP2) method uses tensor hypercontraction (\THC), approximating a electron Coulomb repulsion integrals (ERI) tensor by decomposing into lower order tensors, with sparsity~\cite{Song:2016:thc}.


\subsubsection{Data Mining}

Tensor decompositions have become a standard approach in brain signal analysis due to multiple heterogeneous data sources.
Some recent methods have been surveyed in~\cite{Cichocki:2013:survey,Cao:2015:tensor-app}. 
Electroencephalogram (EEG) and fMRI data are treated as tensors and analyzed by different tensor decompositions (e.g., \CPD) to study the structure of epileptic seizures~\cite{Acar:2007:epilepsy,Acar:2011:missing-data}, better understand the active brain regions and their behavior~\cite{Latchoumane:2012:tensor-app,Davidson:2013:tensor-app}, do feature selection~\cite{Cao:2014:tensor-app}, and model neuroimaging data~\cite{Morup:2008:tensor-app}.
BrainQ is a widely available tensor dataset consisting of a sparse tensor with (subject, brain-voxel, noun) as dimensions and a matrix (noun, properties), which are measured from brain activity where individual subjects are shown nouns. 
Factorizing this is known as a coupled factorization~\cite{Acar:2011:allatonce}, and Papalexakis, et al. demonstrated a scalable method using random sampling~\cite{Papalexakis:2014:turbosmt}.
On the supervised learning setting, F.~Wang et al. used fMRI data and adapted the Sparse Logistic Regression to accept tensor input that consequently avoided the loss of correlation information among different orders~\cite{Wang:2014:tensor-app}.

Personalized web search tailors the results of a search query for a particular user by utilizing the click history of this user's previous search results. 
Researchers constructed tensors from (user, query, webpage) information and used \CPD~\cite{Kolda:2006:tophits} and Tucker decompositions~\cite{Sun:2005:cubesvd} to tackle this problem.

Recommendation systems have also found tensor methods effective to resolve overloaded tags.
Some approaches have been explored using \CPD and Tucker decompositions and their variants on collaborative filtering~\cite{Xu:2006:tensor-app}, a tag-recommendation engine~\cite{Symeonidis:2008:tensor-app,Rendle:2009:tensor-app,Karatzoglou:2010:tensor-app}, personalized tags~\cite{Fang:2014:dtt}, and sparse international relationships~\cite{Schein:2015:tensor-app}.

\subsubsection{Signal Processing}
There has been an extensive research from the Signal Processing community, which examines theoretical aspects of tensor methods~\cite{Jiang:2004:kruskal} such as identifiability, or improves existing decompositions~\cite{Bro:1999:trilinear,Sidiropoulos:2000:blind}. 
A tutorial addressing signal processing applications can be found in~\cite{Cichocki:2015:survey}.
Please refer to the survey~\cite{Sidiropoulos:2017:survey} for more complete applications in signal processing.

\subsubsection{Other Areas}


The usage of tensors and tensor decompositions as tools facilitating the extraction of useful information out of complex data is not limited to the categories mentioned above. 
For example, Benson, et al. used Tucker decomposition to compress scientific data obtained by Direct Numerical Simulation (DNS)~\cite{Austin:2015:tucker}.
Song et al. applied \CPD to forecast of the power demand and detect anomalies in smart electrical grid~\cite{Song:2017:PowerCast}.
A variant of Tucker decomposition was used in AC optimal power flow in the work~\cite{Oh:2016:TensorsIP}.
\TT was used in the hierarchical uncertainty quantification to reduce the computational cost of circuit simulation~\cite{zhang2015enabling}.
Electronic design automation (EDA) problems employed \CPD, Tucker, and \TT decompositions to ease the suffer of the curse of dimensionality~\cite{zhang2017tensor}.
Motion control problems in the context of robotics took \TT into consider for its compressed representations~\cite{gorodetsky2015efficient}.

\section{Benchmark Workloads}  \label{sec:workload}
This section we describe the workloads in \BenchName, which includes element-wise addition/subtraction/ multiplication/division, tensor-scalar, tensor-times-vector, tensor-times-matrix, and tensor-times-matrix sequence operations.
We referred to the surveys~\cite{Kolda:2009:survey,Cichocki:2016:survey,Cichocki:2014:survey,DeLathauwer:2008:block-survey-1,Cichocki:2015:survey,Anandkumar:2014:survey,Sidiropoulos:2017:survey} and papers~\cite{Li:2018:thesis} for these definitions.

A tensor, abstractly defined, is a function of three or more indices.
In computational data analytics, one may regard a tensor as a multidimensional array, where each of its dimensions is also called a \emph{mode} and the number of dimensions or modes is its \emph{order}.
For example, a scalar is a tensor of order 0; a vector is a tensor of order 1; and a matrix, order 2, with two modes (its rows and its columns).
Notationally, we represent tensors as calligraphic capital letters, e.g., $\T{X} \in \R^{I \times J \times K}$;
matrices by boldface capital letters, e.g., $\M{U} \in \R^{I \times J}$;
vectors by boldface lowercase letters, e.g., $\V{x} \in \R^I$;
and scalars by lowercase letters, such as $x_{ijk}$ for the $(i,j,k)$ element of a third-order tensor $\T{X}$.
A \emph{slice} is a two-dimensional cross-section of a tensor, achieved by fixing all mode indices but two, e.g., $\M{S}_{::k} = \T{X}(:,:,k)$ in MATLAB notation. 
A \emph{fiber} is a vector extracted from a tensor along some mode, selected by fixing all indices but one, e.g., $\V{f}_{:jk} = \T{X}(:,j,k)$. 

A tensor can be reshaped to a matrix, which is called matricization. 
For a tensor $\T{X} \in \R^{I_1 \times \dots \times I_n \times \dots \times I_N}$, its matricized tensor along with mode-$n$ is $\M{X}_{(n)} \in \R^{I_1 \cdots I_{n-1} I_{n+1} \cdots I_N \times I_n}$. A matrix can be also reshaped to a tensor by splitting one mode into two or more.

\subsection{Tensor Element-Wise Operations}
\label{sec:tew}
Tensor element-wise (\TEW) operations include addition, subtraction, multiplication, and division operations, which are applied to every corresponding pair of elements from two tensor objects if they have the same order and shape (dimension sizes).
For example, element-wise tensor addition of $\T{X}, \T{Y} \in \R^{I_1 \times \dots \times I_N}$ is $\T{Z} = \T{X} .+ \T{Y}$, where
\begin{equation}
z_{i_1 \cdots i_N} = x_{i_1 \cdots i_N} + y_{i_1 \cdots i_N}.
\end{equation}
Similarly for element-wise tensor subtraction $\T{Z} = \T{X} .- \T{Y}$, multiplication $\T{Z} = \T{X} .* \T{Y}$, and division $\T{Z} = \T{X} ./ \T{Y}$. When the two input tensors have exactly the same non-zero distribution, element-wise operations can be easily implemented by iterating all non-zeros of the two sparse tensors and doing the corresponding operation for each element.
The tricky cases are when the non-zero patterns of tensors \T{X} and \T{Y} are different and even worse they could be in different shapes.
For these two cases, we cannot easily predict the output tensor \T{Z}'s storage space before computation. 
These two cases we use dynamic vectors and an optimization strategy for parallel algorithms.

\subsection{Tensor-Scalar Operations}
\label{sec:ts}
A Tensor-Scalar (\TS) operation is the addition (\TSA) /subtraction (\TSS) /multiplication (\TSM) /division (\TSD) of a tensor $\T{X} \in \R^{I_1 \times I_N}$ with a scalar $s \in \R$ for every non-zero entry. It is denoted by $\T{Y} = \T{X} \times s$.
For example, the \TSM operation is defined as
\begin{equation}
y_{i_1 \cdots i_{n-1} r i_{n+1} \cdots i_N} = s \times x_{i_1 \cdots i_{n-1} i_n i_{n+1} \cdots i_N} .
\end{equation}
Since $\T{Y} = \T{X} \times s$ is the same with $\T{Y} = \T{X} / s^{-1}$ and $\T{Y} = \T{X} + s$ is the same with $\T{Y} = \T{X} - (-s)$, so implementing \TSA and \TSM is enough.

\subsection{Tensor-Times-Vector Operation}
\label{sec:ttv}
The Tensor-Times-Vector (\TTV) in mode $n$ is the multiplication of a tensor $\T{X} \in \R^{I_1 \times \dots \times I_n \times \dots \times I_N}$ with a vector $\V{V} \in \R^{I_n}$, along mode $n$, and is denoted by $\T{Y} = \T{X} \times_n \V{V}$.
This results in a $I_1 \times \dots \times I_{n-1} \times I_{n+1} \times \dots \times I_N$ tensor which has one less dimension.
Its operation is defined as
\begin{equation}
y_{i_1 \cdots i_{n-1} i_{n+1} \cdots i_N} = \sum_{i_n=1}^{I_n} x_{i_1 \cdots i_{n-1} i_n i_{n+1} \cdots i_N} v_{i_n}.
\end{equation}

\subsection{Tensor-Times-Matrix Operation}
\label{sec:ttm}
The Tensor-Times-Matrix (\TTM) in mode $n$, also known as the $n$-mode product, is the multiplication of a tensor $\T{X} \in \R^{I_1 \times \dots \times I_n \times \dots \times I_N}$ with a matrix $\M{U} \in \R^{I_n \times R}$, along mode $n$, and is denoted by $\T{Y} = \T{X} \times_n \M{U}$.\footnote{Our convention for the dimensions of \M{U} differs from that of Kolda and Bader's definition~\cite{Kolda:2009:survey}. In particular, we transpose the matrix modes \M{U}, which leads to a more efficient \TTM under the row-major storage convention of the C language.}
This results in a $I_1 \times \dots \times I_{n-1} \times R \times I_{n+1} \times \dots \times I_N$ tensor, and its operation is defined as
\begin{equation}
\label{eq:ttm}
y_{i_1 \cdots i_{n-1} r i_{n+1} \cdots i_N} = \sum_{i_n=1}^{I_n} x_{i_1 \cdots i_{n-1} i_n i_{n+1} \cdots i_N} u_{i_n r}.
\end{equation}
\TTM is a special case of tensor contraction.
We consider \TTM specifically because of its more common usage in tensor decompositions for data analysis, such as the Tucker decomposition.
Also, note that $R$ is typically much smaller than $I_n$ in such decompositions, and typically $R < 100$.

\TTM is also equivalent to a matrix-matrix multiplication in the following form:
\begin{equation}
  \T{Y} = \T{X} \times_n \M{U} \quad \Leftrightarrow \quad \M{Y}_{(n)} = \M{U} \M{X}_{(n)}.
  \label{eq:moden-mat}
\end{equation}
Therefore, one feasible way to implement an \TTM is to first matricize the tensor, then use an optimized matrix-matrix multiplication to compute the matricized output \T{Y}, and, finally, tensorize to obtain \T{Y}.
However it has the tensor-matrix transformation as the extra overhead and does not work well for sparse tensors.

\subsection{Kronecker and Khatri-Rao Products}
Kronecker and Khatri-Rao products are both matrix products.
The \emph{Kronecker product} generalizes the outer product for matrices.
Given $\M{U} \in \mathbb{R}^{I \times J}$ and $\M{V} \in \mathbb{R}^{K \times L}$, the Kronecker product $\M{U} \otimes \M{V} \in \R^{IK \times JL}$ is 
\begin{equation}
\M{U} \otimes \M{V} =
\begin{bmatrix}
u_{11}\M{V} & u_{12}\M{V} & \cdots & u_{1J}\M{V} \\
u_{21}\M{V} & u_{22}\M{V} & \cdots & u_{2J}\M{V} \\
\vdots & \vdots & \ddots & \vdots \\
u_{I1}\M{V} & u_{I2}\M{V} & \cdots & u_{IJ}\M{V}
\end{bmatrix}
\end{equation}

The \emph{Khatri-Rao product} is a ``matching column-wise'' Kronecker product between two matrices with the same number of columns.
Given matrices $\M{A} \in \R^{I \times R}$ and $\M{B} \in \R^{J \times R}$, their Khatri-Rao product is denoted by $\M{A} \odot \M{B} \in \R^{(IJ) \times R}$,
\begin{equation}
{
\M{A} \odot \M{B} = \left[ \V{a}_1 \otimes \V{b}_1, \V{a}_2 \otimes \V{b}_2, \dots, \V{a}_R \otimes \V{b}_R  \right],
}
\end{equation}
where $\V{a}_r$ and $\V{b}_r$, $r=1,\dots,R$, are columns of $\M{A}$ and $\M{B}$.

Kronecker and Khatri-Rao products appear frequently in tensor decompositions that are formulated as matrix operations.
However, such formulations typically also require redundant computation or extra storage to hold matrix operands, so in practice these operations are tend to be not implemented directly but rather integrated into tensor operations.

\subsection{Tensor-Times-Matrix Sequence Operation}
\label{sec:mttkrp}
There are two types of tensor-times-matrix sequence operations, \TTM chain and \MTTKRP.
\TTM chain is a sequence of \TTM operations with one's output as the next one's input.
An alternative way to think \TTM chain is a matriced tensor times the Kronecker product of matrices.
\MTTKRP, matricized tensor times Khatri-Rao product, is a matricized tensor times the Khatri-Rao product of matrices.
For an $N^{th}$-order tensor \T{X} and given matrices $\M{U}^{(1)}, \ldots, \M{U}^{(N)}$, the mode-$n$ \MTTKRP is
\begin{equation}
\label{eq:mttkrp-init}
\tilde{\M{U}}^{(n)} = \M{X}_{(n)} \left(\odot_{i=1, \dots, N}^{i \ne n} \M{U}_i \right) = \M{X}_{(n)} \left( \M{U}^{(N)} \odot \cdots \odot \M{U}^{(n+1)} \odot \M{U}^{(n-1)} \odot \cdots \odot \M{U}^{(1)} \right),
\end{equation}
where $\M{X}_{(n)}$ is the mode-$n$ matricization of tensor $\T{X}$, $\odot$ is the Khatri-Rao product.

\subsection{Others}
We also provide the transformation between tensors and matrices and some sorting algorithms for sparse tensors.
\section{Data Structures, Algorithms, and Implementations}  \label{sec:impl}

\subsection{Data Structures}
\begin{figure}[h]
  \centering
  \includegraphics[width=0.15\linewidth]{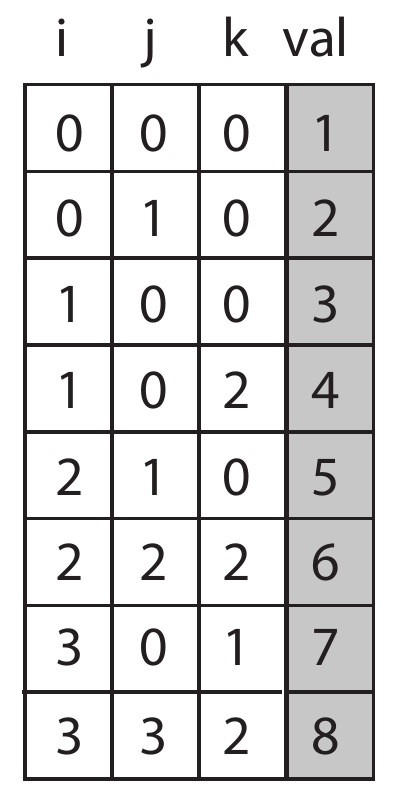}
  \caption{COO format of an example $4 \times 4 \times 3$ tensor.}
  \label{fig:coo}
\end{figure}

Since COO~\cite{Kolda:2009:survey} is the simplest and arguably de facto standard way to store a sparse tensor, and it is mode generic, we only support COO format in this work.
Other state-of-the-art formats will be included as our future work.
We use \inds and \val to represent the indices and values of the non-zeros of a sparse tensor respectively. 
\val is a size-$\nnz$ array of floating-point numbers, \inds is a size-$\nnz$ array of integer tuples. 
Figure~\ref{fig:coo} shows a $4 \times 4 \times 3$ sparse tensor in COO format.
The indices of each mode are represented as $i$, $j$, and $k$. 
Observe that some indices in \inds repeat, for example, entries $(1,0,0)$ and $(1,0,2)$ have the same $i$ and $j$ indices.
This redundancy suggests some compression of this indexing metadata should be possible, as proposed in some work~\cite{Smith:2015:splatt,Liu:2017:fcoo}.

\subsection{Algorithms}
This section describes the sequential algorithms for the workloads in \Cref{sec:workload}.
All algorithms directly operates on the input sparse tensor(s) without explicit tensor-matrix transformation.

\begin{algorithm}[h]
\small
\caption{Sequential COO-\TEW-eq-Addition algorithm for tensors in the same order and shape.}
\begin{algorithmic}[1]
  \Require
       A third-order sparse tensor $\T{X}, \T{Y} \in \R^{I \times J \times K}$ with $\nnz$ non-zeros;
  \Ensure
       Sparse tensor $\T{Z} \in \R^{I \times J \times K}$;

  \Statex \Comment{$\T{Z} = \T{X} .+ \T{Y}$} 

  \State Allocate \T{Z} space with $\nnz$ non-zeros; \Comment{Pre-allocation space.}

  \For{$m=1, \dots, \nnz$}
  	\State $\inds^1_z(m) = \inds^1_x(m)$, $\inds^2_z(m) = \inds^2_x(m)$, $\inds^3_z(m) = \inds^3_x(m)$;
  	\State $\val_z(m) = \val_x(m) + \val_y(m)$;
  \EndFor \\
  \Return{} $\T{Z}$;
\end{algorithmic}
\label{alg:tew-eq}
\end{algorithm}

\subsubsection{\TEW}
As mentioned in \Cref{sec:tew}, \TEW operation has two cases: one is between two tensors in exactly the same shape and non-zero distribution; the other only requires the two tensors are in the same tensor order. 

For the first case, we show \TEW addition as an example in \Cref{alg:tew-eq}. 
The output tensor has the same shape and non-zero distribution with the two input tensors, thus it can be pre-allocated.
Then the calculation simply does addition by looping all non-zeros.

\begin{algorithm}[h]
\small
\caption{Sequential COO-\TEW-Addition algorithm for general tensors.}
\begin{algorithmic}[1]
  \Require
       A third-order sparse tensor $\T{X} \in \R^{I_1 \times J_1 \times K_1}$ with $\nnz_1$ non-zeros, $\T{Y} \in \R^{I_2 \times J_2 \times K_2}$ with $\nnz_2$ non-zeros;
  \Ensure
       Sparse tensor $\T{Z} \in \R^{I_3 \times J_3 \times K_3}$;

  \Statex \Comment{$\T{Z} = \T{X} .+ \T{Y}$} 

  \State $I_3 = max\{I_1, I_2\}$, $J_3 = max\{J_1, J_2\}$, $K_3 = max\{K_1, K_2\}$ \Comment{Unify the tensor shape.}

  \State Sort \T{X} and \T{Y} in the same dimension order.

  \State $m_1=1$, $m_2=1$
  \While{$m_1<\nnz_1$ and $m_2<\nnz_2$}
  	\If{$\inds_x == \inds_y$}
  		\State Append($\inds^1_z$, $\inds^1_x(m_1)$); Append($\inds^2_z$, $\inds^2_x(m_1)$); Append($\inds^3_z$, $\inds^3_x(m_1)$);
  		\State Append($\val_z$, $\val_x(m_1) + \val_y(m_2)$);
	\EndIf
  	\If{$\inds_x > \inds_y$}
  		\State Append($\inds^1_z$, $\inds^1_y(m_1)$); Append($\inds^2_z$, $\inds^2_y(m_1)$); Append($\inds^3_z$, $\inds^3_y(m_1)$);
  		\State Append($\val_z$, $\val_y(m_2)$);
  	\EndIf
  	\If{$\inds_x < \inds_y$}
  		\State Append($\inds^1_z$, $\inds^1_x(m_1)$); Append($\inds^2_z$, $\inds^2_x(m_1)$); Append($\inds^3_z$, $\inds^3_x(m_1)$);
  		\State Append($\val_z$, $\val_x(m_1)$);
  	\EndIf
  \EndWhile 
  \If{$m_1 < \nnz_1$}
  	\State Append($\inds^1_z$, $\inds^1_x(m_1,:)$); Append($\inds^2_z$, $\inds^2_x(m_1,:)$); Append($\inds^3_z$, $\inds^3_x(m_1,:)$);
  	\State Append($\val_z$, $\val_x(m_1,:)$)
  \EndIf
  \If{$m_2 < \nnz_2$}
  	\State Append($\inds^1_z$, $\inds^1_x(m_2,:)$); Append($\inds^2_z$, $\inds^2_x(m_2,:)$); Append($\inds^3_z$, $\inds^3_x(m_2,:)$);
  	\State Append($\val_z$, $\val_x(m_2,:)$)
  \EndIf
  \\
  \Return{} $\T{Z}$;
\end{algorithmic}
\label{alg:tew}
\end{algorithm}

For the second case, its algorithm is shown in \Cref{alg:tew}.
The output tensor size is set by the maximum dimension size of the two input tensors.
Since we do not know the number of the output non-zeros, we cannot pre-allocate the space of the output tensor \T{Z} but using dynamic allocation to append non-zeros.
First, we need to sort tensors \T{X} and \T{Y} in the order of mode $1 \succ 2 \succ 3$, then compare the indices in lexicographical order for each non-zero pair-to-pair, e.g., indices $(2,1,1) > (1,1,2) > (1,1,1)$.
If two indices are the equal, then we append the indices and the sum of the two non-zero values to the output \T{Z}.
Otherwise, we append the smaller indices and its corresponding value to \T{Z}.
Only if we run out of non-zeros in either \T{X} or \T{Y}, we append the rest indices and values of the other one to \T{Z}.

\begin{algorithm}[h]
\small
\caption{Sequential COO-\TSM algorithm.}
\begin{algorithmic}[1]
  \Require
       A third-order sparse tensor $\T{X} \in \R^{I \times J \times K}$ with $\nnz$ non-zeros;
  \Ensure
       Output sparse tensor $\T{Y} \in \R^{I \times J \times K}$;

  \Statex \Comment{$\T{Y} = \T{X} \times s$} 

  \State Allocate \T{Y} space with $\nnz$ non-zeros; \Comment{Pre-allocation space.}

  \For{$m=1, \dots, \nnz$}
  	\State $\inds^1_y(m) = \inds^1_x(m)$, $\inds^2_y(m) = \inds^2_x(m)$, $\inds^3_y(m) = \inds^3_x(m)$;
  	\State $\val_y(m) = s \times \val_x(m)$;
  \EndFor \\
  \Return{} $\T{Y}$;
\end{algorithmic}
\label{alg:ts}
\end{algorithm}

\subsubsection{\TS}
\TS algorithm is simple. The output \T{Y} can be pre-allocated and computed by looping all non-zeros.
\Cref{alg:ts} shows the \TSM algorithm.

\begin{algorithm}[h]
\small
\caption{Sequential COO-\TTV algorithm.}
\begin{algorithmic}[1]
  \Require
       A third-order sparse tensor $\T{X} \in \R^{I \times J \times K}$, dense vector $\M{V} \in \R^{K}$, mode $n=3$;
  \Ensure
       Sparse tensor $\T{Y} \in \R^{I \times J}$;

  \Statex \Comment{$\T{Y} = \T{X} \times_n \V{V}$} 
  \State Pre-process to obtain $\nfibs$: the number of mode-n fibers of \T{X} and $\fptr$: the beginnings of each \T{X} mode-n fiber, sized $\nfibs$.
  \State Allocate \T{Y} space with $\nfibs$ non-zeros; \Comment{Pre-allocation space.}

  \For{$f=1, \dots, \nfibs$}
    \State $\inds^1_Y(f) = \inds^1_X(\fptr(f))$, $\inds^2_Y(f) = \inds^2_X(\fptr(f))$
    \For{$m=\fptr(f), \dots, \fptr(f+1)-1$}
      \State $k = \inds^3_X(m)$
      \State $\val_Y(f) += \val_X(m) \times u(k)$
    \EndFor
  \EndFor
  \State \textbf{Return} \T{Y};
\end{algorithmic}
\label{alg:ttv}
\end{algorithm}

\subsubsection{\TTV}
\TTV algorithm in mode-$n$ is shown in \Cref{alg:ttv}. 
It first pre-compute the number of fibers $\nnz_F$ of input tensor \T{X} and the beginning positions of each fiber.
Then we can pre-allocate the output tensor \T{Y} with $\nnz_F$, because this product does not influence the non-zero layout for $I$ and $J$ modes.
The algorithm loops all the fibers of \T{X}, and a reduction happens for all non-zeros in each fiber.

\begin{algorithm}[ht]
\small
\caption{Sequential COO-\TTM algorithm~\cite{Li:2016:spttm}.}
\begin{algorithmic}[1]
  \Require
       A sparse tensor $\T{X} \in \R^{I \times J \times K}$, a dense matrix $\M{U} \in \R^{K \times R}$, and an integer $n = 3$;
  \Ensure
       Sparse tensor $\T{Y} \in \R^{I \times J \times R}$;
  \Statex \Comment{$\T{Y} = \T{X} \times_n \M{U}$}

  \State Pre-process to obtain $\nfibs$: the number of mode-n fibers of \T{X} and $\fptr$: the beginnings of each \T{X} mode-n fiber, size $\nfibs$.
  \State Allocate \T{Y} space with $\nfibs \times R$ non-zeros; \Comment{Pre-allocation space.}

  \For{$f=1, \dots, \nfibs$}
    \State $i = \inds^1_X(\fptr(f))$, $j = \inds^2_X(\fptr(f))$
    \For{$r=1, \dots, R$}
      \State $\inds^1_Y(f \times R + r) = i$, $\inds^2_Y(f \times R + r) = j$, $\inds^3_Y(f \times R + r) = r$
    \EndFor
    \For{$m=\fptr(f), \dots, \fptr(f+1)-1$}
      \State $k = \inds^3_X(m)$
      \State $value = \val_X(m)$
      \For{$r=1, \dots, R$}
        \State $\val_Y(f \times R + r) += value \times u(k \times R + r)$
      \EndFor
    \EndFor
  \EndFor

  \State \textbf{Return} \T{Y};
\end{algorithmic}
\label{alg:ttm}
\end{algorithm}

\subsubsection{\TTM}
\TTM algorithm is illustrated in \Cref{alg:ttm}.
Similarly to \TTV algorithm, we obtain the number of fibers $\nnz_F$ and the beginning positions of each fiber then $\nnz_F \times R$ space are allocated for the output tensor \T{Y}.
The algorithm loops all the $\nnz_F$ fibers and does a reduction between sized-$R$ vectors.
This \TTM algorithm directly operates on the input sparse tensor by avoiding tensor transformation.
The explanation of \Cref{alg:ttm} can be found in the work~\cite{Li:2016:spttm,Ma:2018:sptucker-gpu}.

\begin{algorithm}[h]
\small
\caption{Sequential COO-\MTTKRP algorithm (\cite{Bader:2007:tensortoolbox-sparse}).}
\begin{algorithmic}[1]
  \Require
       A third-order sparse tensor $\T{X} \in \R^{I \times J \times K}$, dense matrices $\M{B} \in \R^{J \times R}, \M{C} \in \R^{K \times R}$;
  \Ensure
       Updated dense matrix $\tilde{\M{A}} \in \R^{I \times R}$;

  \Statex \Comment{$\tilde{\M{A}} \leftarrow \T{X}_{(1)} (\M{C} \odot \M{B})$} 

  \For{$m=1, \dots, \nnz$}
  	\State $i = \inds^1(m)$, $j = \inds^2(m)$, $k = \inds^3(m)$;
    \State $value = \val(m)$
  	\For{$r=1, \dots, R$}
  		\State  $\tilde{A}(i \times R + r) += value \times C(k \times R + r) \times B(j \times R + r)$
  	\EndFor
  \EndFor \\
  \Return{} $\tilde{\M{A}}$;
\end{algorithmic}
\label{alg:mttkrp}
\end{algorithm}

\subsubsection{\MTTKRP}
\MTTKRP algorithm is shown in \Cref{alg:mttkrp}, the output matrix of which is initialized before and only needs to be updated.
This algorithm loops all non-zeros of the tensor \T{X} and times the corresponding two matrix vectors, to update the designated output matrix vector.
Readers can refer more details of this algorithm in~\cite{Bader:2007:tensortoolbox-sparse}.

\begin{table*}[htbp]
\centering
{
\caption{The analysis of data storage and their algorithms for third-order cubical tensors ($\T{X} \in \R^{I \times I \times I}$). We consider all input tensors with \nnz nonzero entries and $\nnz_F$ fibers, $I \ll \nfibs \ll \nnz$. The indices use $32$ bits, and values are single-precision floating-point numbers with $32$ bits.}
\label{tab:analysis}
\begin{tabular}{rcccc}
  \toprule
  
  \multirow{2}{*}{Workloads} & 
  Storage & 
  Work & 
  Memory & 
  Arithmetic \\

  & 
  (Bytes) &
  (Flops) &
  Access (Bytes) &
  Intensity (AI) \\ \hline

  \TEW & 
  $48 \nnz $ & 
  $\nnz$ & 
  $36 \nnz$ &
  $1/36$ \\

  \TS & 
  $32 \nnz $ & 
  $\nnz$ & 
  $32 \nnz$ &
  $1/32$ \\

  \TTV & 
  $(16 \nnz + 12 \nfibs) $ & 
  $2 \nnz$ & 
  $(12 \nnz + 20 \nfibs)$ &
  $\sim 1/6$ \\

  \TTM & 
  $(16 \nnz + 16 \nfibs R + 4IR) $ & 
  $2 \nnz R$ & 
  $4 \nnz R + 8 \nnz + 12 \nfibs R + 8 \nfibs$ &
  $\sim 1/2$ \\

  \MTTKRP & 
  $(16 \nnz + 12 IR)$ & 
  $3 \nnz R$ & 
  $12 \nnz R + 16 \nnz$ &
  $\sim 1/4$ \\

  \bottomrule
\end{tabular}
}
\end{table*}

According to the above algorithms, we compute the storage, the number of floating-point operations (Flops), the amount of memory access in bytes, and the arithmetic intensity (the ratio of \#Flops/\#Bytes) in \Cref{tab:analysis}.
For simplicity, we use a cubical third-order sparse tensor $\T{X} \in \R^{I \times I \times I}$ with $\nnz$ non-zeros and $\nnz_F$ fibers as an example.
Because of the irregular access pattern of sparse tensors, the memory access does not consider the cache effect.
All workloads have arithmetic intensity less than $1$, thus it is hard to easily achieve good performance on common architectures.
While \MTTKRP has the most Flops and memory access, its arithmetic intensity is smaller than \TTM, which it $\sim 1/2$.
\TEW and \TS have the smallest arithmetic intensity and the largest storage due to the output tensor.
Despite of different algorithm behavior, these algorithms are generally considered memory intensive, which demonstrates the emphasis of our \BenchName.

\subsection{Multicore Implementations}

Some workloads are easy to parallelize. 
We parallelize the loop of all non-zeros in \TEW-eq (\Cref{alg:tew-eq}) and \TS (\Cref{alg:ts}).
For \TTV (\Cref{alg:ttv}) and \TTM (\Cref{alg:ttm}), the loop of fibers is parallelized because each fiber computation is independent.

\TEW (\Cref{alg:tew}) is difficult to be parallelized because of its dynamic append operations and no pre-allocation available.
We partition the two tensors in such a way that there is no overlap between their indices, then we run \TEW algorithm locally for a sub-tensor in each thread and append the results to a local output buffer.
The partitioning first split one of the two tensors (say \T{X}) by slices and meanwhile tend to evenly distribute its non-zeros. 
This makes sure that all non-zeros of a slice cannot be split into two partitions.
Then the partitioning of the other tensor (say \T{Y}) is according to this slice partitioning strategy.
In this case, we assure every partition does not overlap with each other, thus they can independently computed in parallel.

We parallelize the loop of all non-zeros of \MTTKRP (\Cref{alg:mttkrp}) as well, but Line 4 may have data race by writing into the same location of $\tilde{\M{A}}$.
We implemented two solutions: 1) Use atomics to protect the correctness, but the performance suffers much; 2) Employ privatization approach to allocate a thread-local buffer. The data is first written to this buffer by each thread privately, then a global reduction for the buffers is used to get the final results. In this case, we can generally get better performance than using atomics.

For these parallel implementations, we have not considered the NUMA effect, which will be another piece of our future work.

\section{Dataset}  \label{sec:dataset}
\BenchName now only considers real-world data as input.
The sparse tensors derived from real-world applications, that appear in \Cref{tab:dataset}, ordered by decreasing non-zero density separately for third- and fourth-order tensors.
Most of these tensors are included in The Formidable Repository of Open Sparse Tensors and Tools (FROSTT) dataset (Refer to the details in \cite{Smith:2017:frostt-dataset}).
The \tennm{darpa} (source IP-destination IP-time triples), \tennm{fb-m}, and \tennm{fb-s} (short for ``freebase-music'' and ``freebase-sampled'', entity-entity-relation triples) are from the dataset of HaTen2~\cite{Jeon:2015:haten2-pak}, and \tennm{choa} is built from electronic health records (EHRs) of pediatric patients at Children's Healthcare of Atlanta (CHOA) \cite{Perros:2017:spartan}.

\begin{table}[htbp]
\centering
\small
{
\caption{Description of sparse tensors.}
\label{tab:dataset}
\begin{tabular}{rcccc}
  Tensors & Order & Dimensions & \#Non-zeros & Density \\
  \toprule
  \tennm{vast} & 3 & $165K \times 11K \times 2$ & 26M & $6.9 \times 10^{-3}$ \\
  \tennm{nell2} & 3 & $12K \times 9K \times 29K$ & 77M & $2.4 \times 10^{-5}$ \\
  \tennm{choa} & 3 & $712K \times 10K \times 767$ & 27M & $5.0 \times 10^{-6}$ \\
  \tennm{darpa} & 3 & $22K \times 22K \times 24M$ & 28M & $2.4 \times 10^{-9}$ \\
  \tennm{fb-m} & 3 & $23M \times 23M \times 166$ & 100M & $1.1 \times 10^{-9}$ \\
  \tennm{fb-s} & 3 & $39M \times 39M \times 532$ & 140M & $1.7 \times 10^{-10}$ \\
  \tennm{deli} & 3 & $533K \times 17M \times 2.5M$ & 140M & $6.1 \times 10^{-12}$ \\
  \tennm{nell1} & 3 & $2.9M \times 2.1M \times 25M$ & 144M & $9.1 \times 10^{-13}$ \\ 
  \midrule
  \tennm{crime} & 4 & $6K \times 24 \times 77 \times 32$ & 5M & $1.5 \times 10^{-2}$ \\
  \tennm{nips} & 4 & $2K \times 3K \times 14K \times 17$ & 3M & $1.8 \times 10^{-6}$ \\ 
  \tennm{enron} & 4 & $6K \times 6K \times 244K \times 1K$ & 54M & $5.5 \times 10^{-9}$ \\ 
  \tennm{flickr4d} & 4 & $320K \times 28M \times 1.6M \times 731$ & 113M & $1.1 \times 10^{-14}$ \\ 
  \tennm{deli4d} & 4 & $533K \times 17M \times 2.5M \times 1K$ & 140M & $4.3 \times 10^{-15}$ \\ 
  \bottomrule
\end{tabular}
}
\end{table}


\section{Experiments}  \label{sec:exp}

We tested these schemes experimentally on a Linux-based Intel Xeon E5-2698 v3 multicore server platform with 32 physical cores distributed on two sockets, each with 2.3\,GHz frequency.
The processor microarchitecture is Haswell, having 32\,KiB L1 data cache and 128\,GiB memory.
The code artifact is written in the C language using OpenMP parallelization, and was compiled using icc 18.0.1.
All experiments use $32$ threads for parallel code except being pointed out otherwise.
The execution time are all averaged by five runs.
For \TTM and \MTTKRP, we set the rank $R = 16$.

We demonstrate the sequential and multicore parallel performance for every workload on the dataset (\Cref{tab:dataset}).

\subsection{\TEW}
\begin{figure}[h]
  \centering
  \footnotesize
  \includegraphics[width=0.55\linewidth]{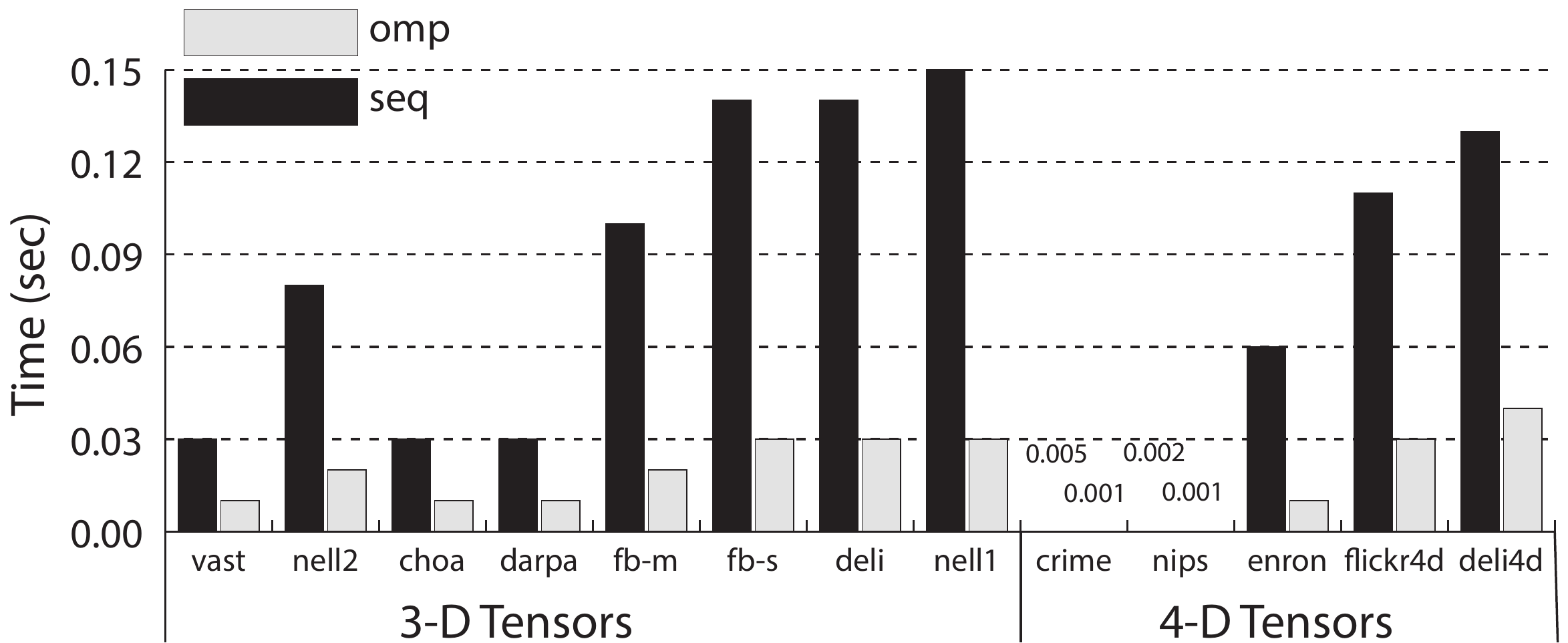} 
  \caption{\TEW-eq-addition for sparse tensors in the same shape and non-zero pattern.}
  \label{fig:tew-eq}
\end{figure}

\begin{figure}[h]
  \centering
  \footnotesize
  \includegraphics[width=0.55\linewidth]{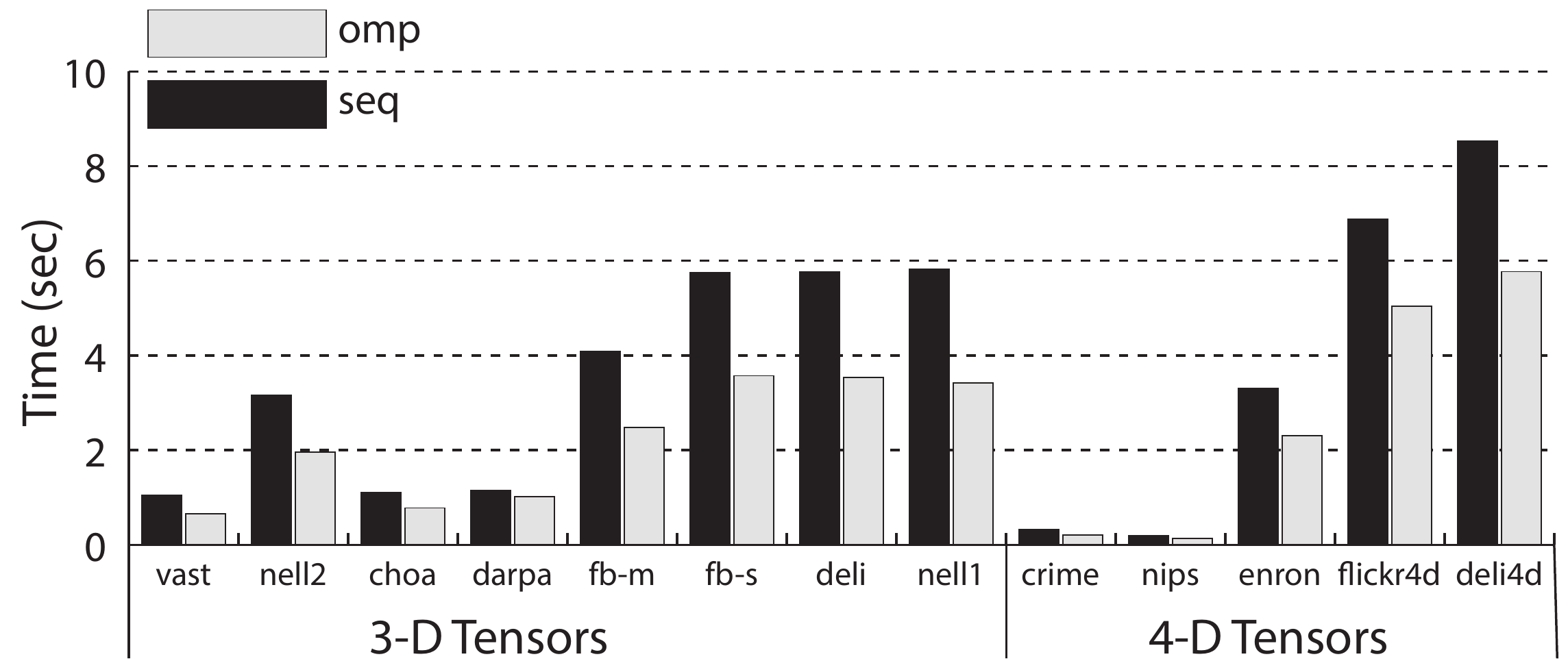} 
  \caption{\TEW-addition for sparse tensors in the same order.}
  \label{fig:tew}
\end{figure}

\Cref{fig:tew-eq} and \ref{fig:tew} show the execution time of the two cases of \TEW addition (\Cref{alg:tew-eq} and \ref{alg:tew}): in the same non-zero pattern and only in the same tensor order, on all third- and fourth-order tensors.
We use the same tensor for the two input for \TEW-eq and \TEW to better show the algorithm effect.
We observe for both cases, parallel \TEW outperforms sequential \TEW.
However, the speedup of \TEW-eq is $3.64 - 5.18 \times$, while the speedup of \TEW is much smaller, which is $1.13 - 1.70 \times$.
This is because: 1) the parallel strategy of \TEW could have a lot more load imbalance than \TEW-eq's even non-zero parallelization; 2) some tensors cannot fully use all $32$ threads due to the slice partitioning (a heavy slice cannot be further partitioned in \Cref{alg:tew}).
Besides, due to the dynamic append operation, the sequential \TEW is tens of times slower than sequential \TEW-eq. 
From our experiments, \TEW subtraction, multiplication, and division behave very similar to \TEW addition in execution time.

\subsection{\TS}
\begin{figure}[h]
  \centering
  \footnotesize
  \includegraphics[width=0.55\linewidth]{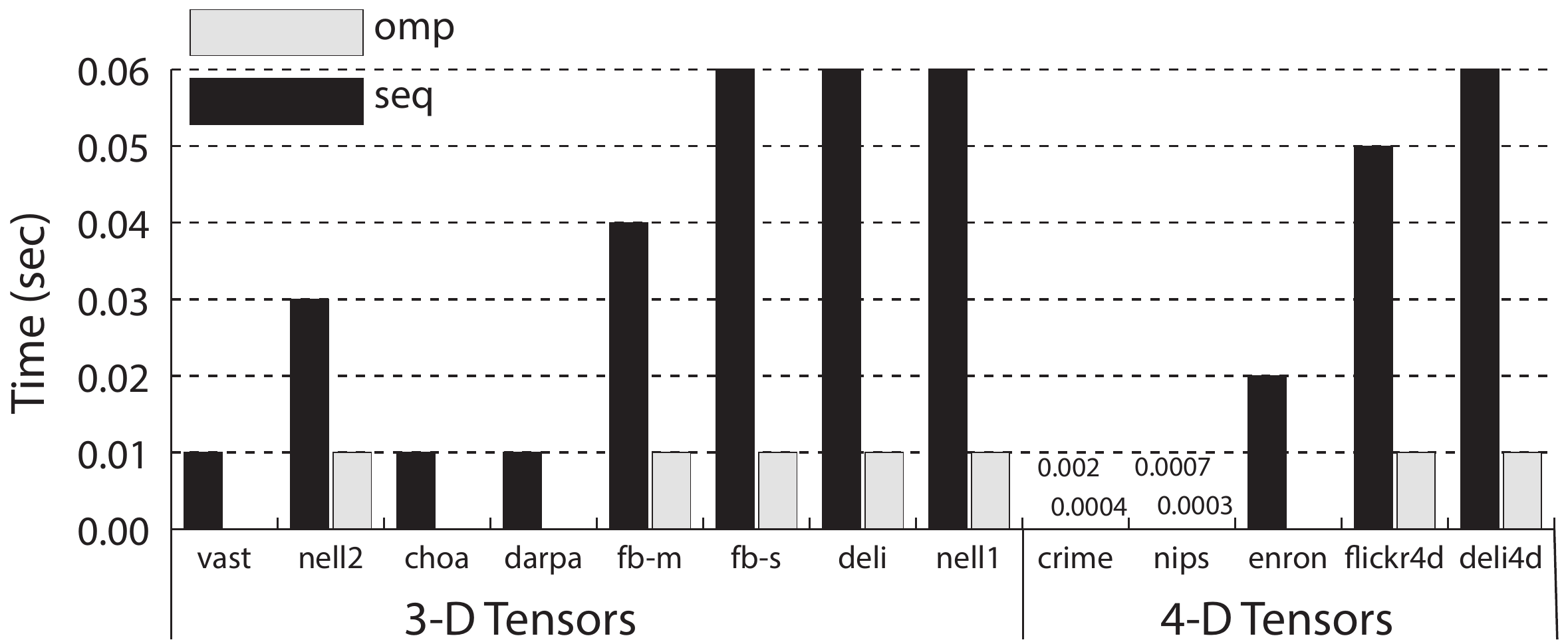} 
  \caption{\TSM execution time.}
  \label{fig:tsm}
\end{figure}

\Cref{fig:tsm} plots the sequential and parallel execution time of \TSM.
Parallel \TSM achieves $2.17 - 5.92 \times$ speedup over sequential \TSM, this is comparable to \TEW-eq in \Cref{fig:tew-eq}.
The sequential \TSM executes faster than the sequential \TEW, which verifies the analysis in \Cref{tab:analysis} and that these two algorithms are memory-bound.
(Because they have the same \#Flops, compute-bound algorithms should have similar execution time.)
From the experiments, the execution times of sequential and parallel \TSA are very close to \TSM.

\subsection{\TTV}
\begin{figure}[h]
  \centering
  \footnotesize
  \includegraphics[width=0.55\linewidth]{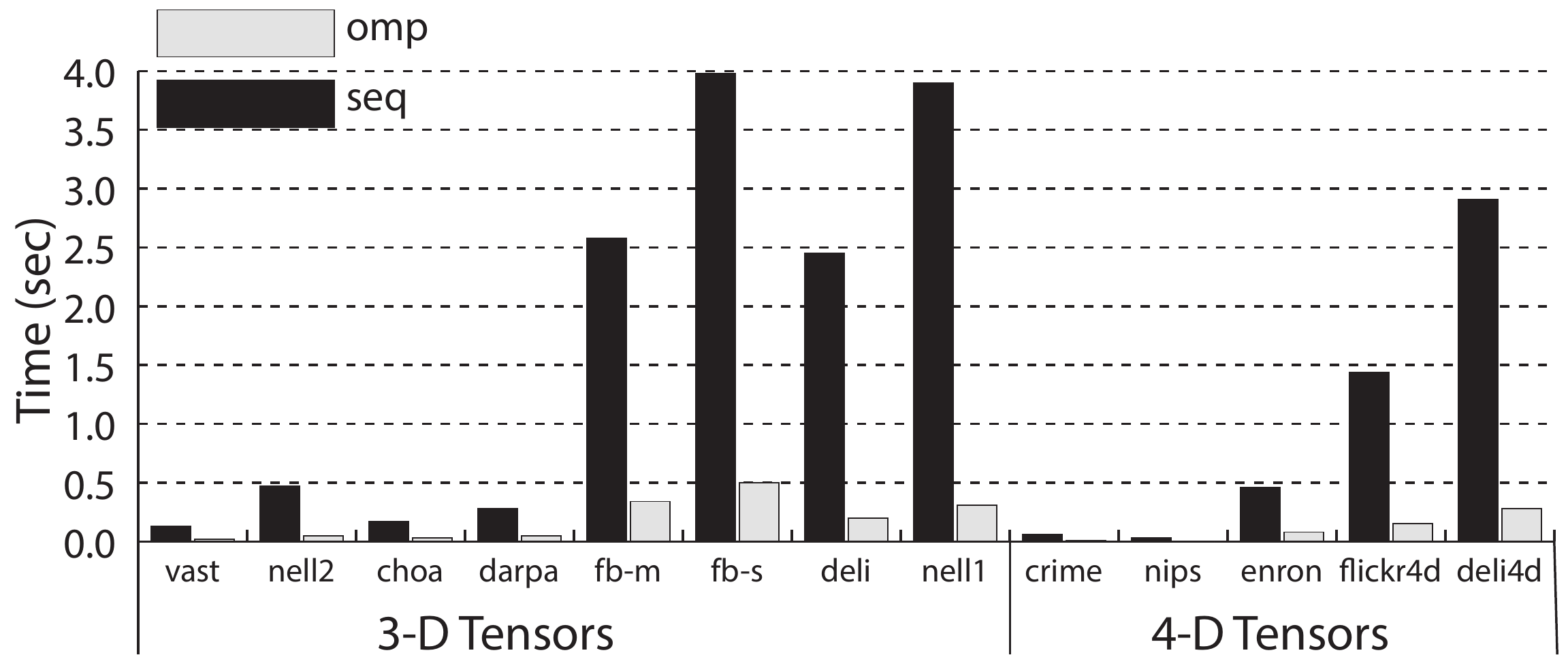} 
  \caption{\TTV: the sum of execution time of all the modes.}
  \label{fig:ttv}
\end{figure}

We illustrate sequential and parallel \TTV time in \Cref{fig:ttv}.
Parallel \TTV outperforms sequential case by $5.21 - 12.45 \times$, this is much higher than the speedup of \TEW-eq, \TEW, and \TSM.
This behavior again matches the analysis in \Cref{tab:analysis} that \TTV has higher arithmetic intensity.
Since higher arithmetic intensity potentially generates less memory contention, thus multicore parallelization could benefit more.

\subsection{\TTM}
\begin{figure}[h]
  \centering
  \footnotesize
  \includegraphics[width=0.55\linewidth]{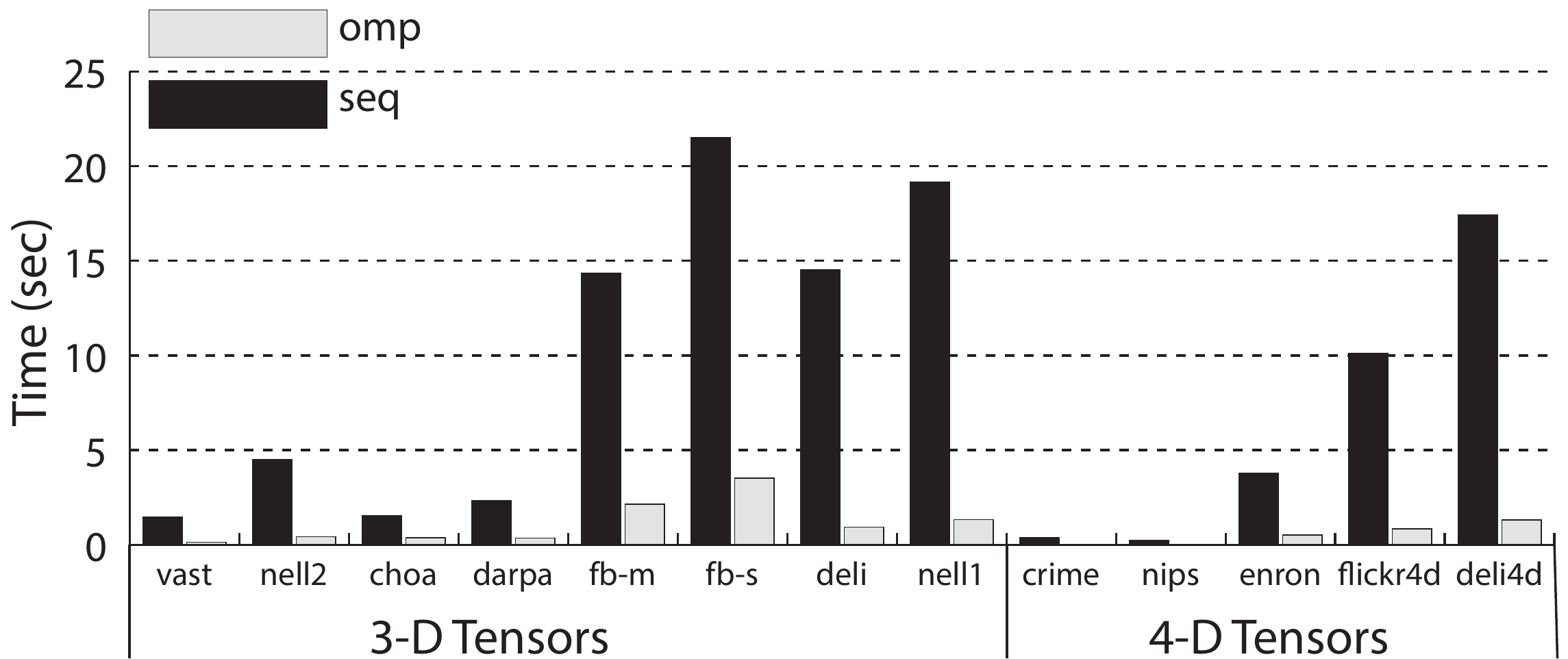} 
  \caption{\TTM: the sum of execution time of all the modes.}
  \label{fig:ttm}
\end{figure}

\Cref{fig:ttm} shows the sequential and parallel execution time of \TTM. 
The speedup of parallel \TTM over sequential case is $4.09 - 15.67 \times$ which is comparable with \TTV's.
This also verifies the analysis that \TTM has the highest arithmetic intensity.
Sequential \TTM is $4.91 - 11.11 \times$ slower than sequential \TTV, that shows the different behavior of timing a dense vector versus a dense matrix.

\subsection{\MTTKRP}
\begin{figure}[h]
  \centering
  \footnotesize
  \includegraphics[width=0.55\linewidth]{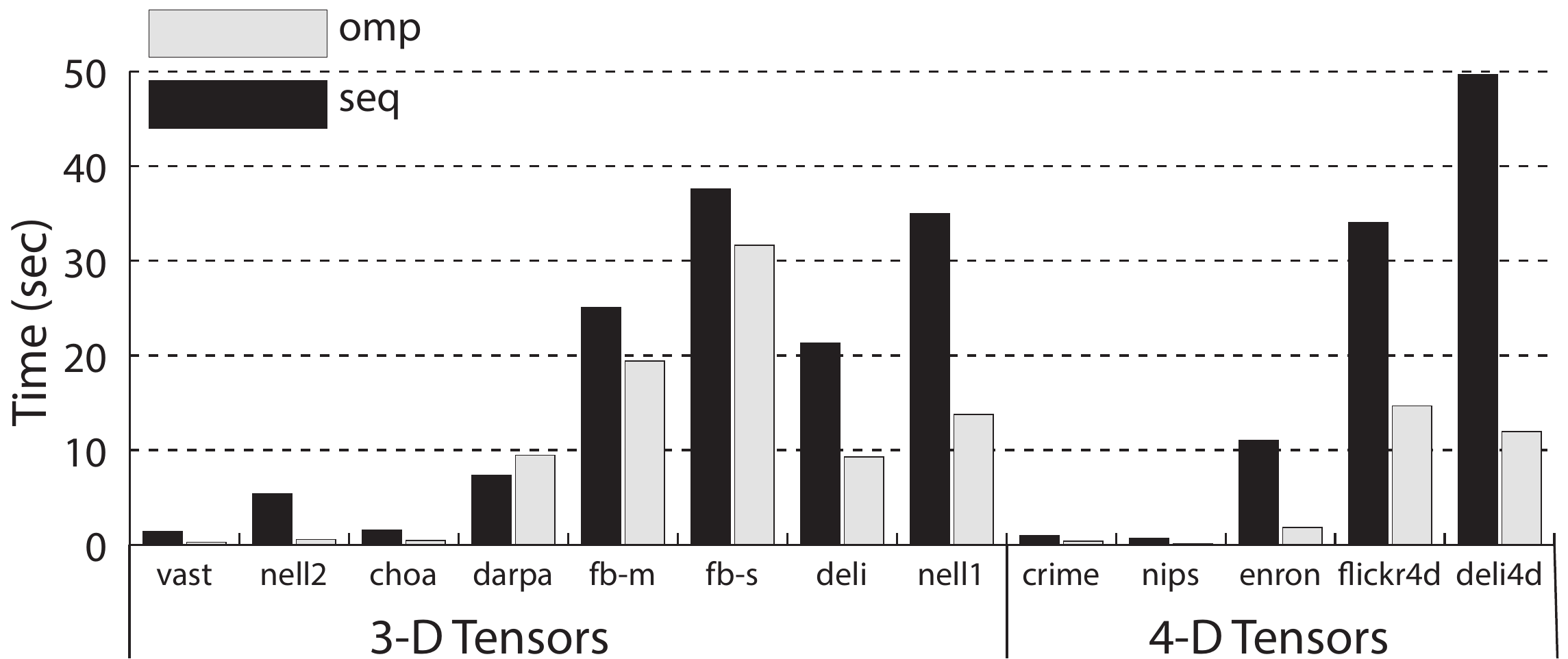} 
  \caption{\MTTKRP: the sum of execution time of all the modes.}
  \label{fig:mttkrp}
\end{figure}

We use privatization technique for parallel \MTTKRP, because it performs better than atomics technique on most of tensors.
The execution time of sequential and parallel \MTTKRP is shown in \Cref{fig:mttkrp}, where the parallel case gains $0.77 - 9.49 \times$ speedup.
For tensor \tennm{darpa}, the only case parallel \MTTKRP is slower than sequential one because of its large thread-local buffer which consumes a large portion of time to do reduction.
The atomics parallel approach could be better in this case, $7.93$ versus $7.32$ (sequential \MTTKRP), but there is still not speedup for this tensor.
\MTTKRP obtains smaller speedup than \TTM and \TTV mainly because data race exists in the output.
Even we use privatization technique to avoid the data race, the extra reduction still take nontrivial amount of time.

From our experiments and analysis above, these relatively simple workloads can well reflect some architecture characteristics. 
This can help architecture designers and application users to evaluate computer systems.

\section{Conclusion}  \label{sec:con}
This work presents a sparse tensor algorithm benchmark suite (\BenchName) for single-core and multi-core CPUs, which is the first sparse tensor benchmark to the best of our knowledge.
\BenchName consists of \TEW, \TS, \TTV, \TTM, \MTTKRP workloads to represent sparse tensor algorithms from different tensor methods in a various application scenarios.
Besides, these workloads can reflect computer architecture features differently from our analysis.

As a benchmark suite, \BenchName already processes good properties such as application and machine diversity, state-of-the-art data structures, algorithms, and optimization techniques included, compatibility for research support, and real-world data set.
Some future work should be done to make \BenchName more complete and robust:
1) more computer systems support, such as GPUs, FPGAs, and distributed systems;
2) more workloads especially tensor-times-tensor product (\TTT);
3) more state-of-the-art sparse tensor formats, e.g., hierarchical COO (HiCOO) and compressed sparse fiber (CSF) format;
4) synthetic data generation for more precise machine performance measurement.


%
\begin{acks}
This research was partially funded by the US Department of Energy, Office for Advanced Scientific Computing (ASCR) under Award No. 66150: "CENATE:  The Center for Advanced Technology Evaluation". Pacific Northwest National Laboratory (PNNL) is a multiprogram national laboratory operated for DOE by Battelle Memorial Institute under Contract DE-AC05-76RL01830.
\end{acks}

%
\bibliographystyle{ACM-Reference-Format}
\bibliography{pasta}

%
\appendix

\end{document}